\documentclass[10pt,compsoc,cspaper, fleqn]{IEEEtran}
\pdfoutput=1
\usepackage{cite}

\ifCLASSINFOpdf
\else
\fi
\usepackage{amsfonts,amssymb,amsmath}
\usepackage{float}
\usepackage{subfigure}
\usepackage{color}
\usepackage{graphicx}
\usepackage{multirow}
\usepackage{alltt}
\definecolor{tabgrey}{rgb}{0.8,0.8,0.8}
\usepackage[mathscr]{eucal}
\usepackage{arydshln}
\usepackage{stmaryrd}
\usepackage{algorithmicx}
\usepackage{algpseudocode}

\usepackage[table,usenames,dvipsnames]{xcolor}
\usepackage[T1]{fontenc}
\usepackage[tableposition=top]{caption}
\usepackage{rotating}
\usepackage{color}
\usepackage{booktabs,array,dcolumn}
\usepackage{multirow}
\usepackage{varwidth}
\usepackage{xcolor,colortbl}
\usepackage{footmisc}

\definecolor{LightCyan}{rgb}{0.88,1,1}
\definecolor{LightGray}{gray}{0.9}
\newcolumntype{a}{>{\columncolor{LightCyan}}r}
\newcolumntype{b}{>{\columncolor[gray]{.9}[5pt]}r|}
\usepackage{hhline}

\newcommand{\ie}{{\sl i.e.}}

\newcommand{\nop}[1]{}

\newcommand{\hide}[1]{\hspace*{-5pt}~}
\floatstyle{ruled}
\newfloat{algorithm}{thp}{loa}
\floatname{algorithm}{Algorithm}

\usepackage{caption}
\usepackage{paralist}

\usepackage{footnote}
\makesavenoteenv{tabular}
\makesavenoteenv{table}

\usepackage{url}

\usepackage{balance}

\begin{document}
\title{A shared latent space matrix factorisation method for recommending new trial evidence for systematic review updates}

\author{Didi Surian\textsuperscript{1}$^\ddagger$, 
		Adam G. Dunn\textsuperscript{1},
		Liat Orenstein\textsuperscript{2},
		Rabia Bashir\textsuperscript{1},\\
		Enrico Coiera\textsuperscript{1},
		and Florence T. Bourgeois\textsuperscript{2,3}
\thanks{1. Centre for Health Informatics, Australian Institute of Health Innovation, Macquarie University, Sydney, Australia.}
\thanks{2. Computational Health Informatics Program, Boston Children's Hospital, Boston, United States.}
\thanks{3. Department of Pediatrics, Harvard Medical School, Boston, United States.}
\thanks{$^\ddagger$Contact author: didi.surian@mq.edu.au}
\thanks{Journal of Biomedical Informatics 79 (2018), p32-40. DOI: 10.1016/j.jbi.2018.01.008}
}

\IEEEtitleabstractindextext{%
\begin{abstract}
\newline
\textbf{Background:} Clinical trial registries can be used to monitor the production of trial evidence and signal when systematic reviews become out of date. However, this use has been limited to date due to the extensive manual review required to search for and screen relevant trial registrations. Our aim was to evaluate a new method that could partially automate the identification of trial registrations that may be relevant for systematic review updates.

\textbf{Materials and Methods:} We identified 179 systematic reviews of drug interventions for type 2 diabetes, which included 537 clinical trials that had registrations in ClinicalTrials.gov. Text from the trial registrations were used as features directly, or transformed using Latent Dirichlet Allocation (LDA) or Principal Component Analysis (PCA). We tested a novel matrix factorisation approach that uses a shared latent space to learn how to rank relevant trial registrations for each systematic review, comparing the performance to document similarity to rank relevant trial registrations. The two approaches were tested on a holdout set of the newest trials from the set of type 2 diabetes systematic reviews and an unseen set of 141 clinical trial registrations from 17 updated systematic reviews published in the Cochrane Database of Systematic Reviews. The performance was measured by the number of relevant registrations found after examining 100 candidates (recall@100) and the median rank of relevant registrations in the ranked candidate lists.

\textbf{Results:} The matrix factorisation approach outperformed the document similarity approach with a median rank of 59 (of 128,392 candidate registrations in ClinicalTrials.gov) and recall@100 of 60.9\% using LDA feature representation, compared to a median rank of 138 and recall@100 of 42.8\% in the document similarity baseline. In the second set of systematic reviews and their updates, the highest performing approach used document similarity and gave a median rank of 67 (recall@100 of 62.9\%).

\textbf{Conclusions:} A shared latent space matrix factorisation method was useful for ranking trial registrations to reduce the manual workload associated with finding relevant trials for systematic review updates. The results suggest that the approach could be used as part of a semi-automated pipeline for monitoring potentially new evidence for inclusion in a review update.
\end{abstract}

\begin{IEEEkeywords}
Systematic reviews, clinical trial, information retrieval, matrix factorisation.
\end{IEEEkeywords}}

\maketitle

\IEEEdisplaynontitleabstractindextext

\IEEEpeerreviewmaketitle

\IEEEraisesectionheading{\section{Background}\label{sec.intro}}

\IEEEPARstart{S}{ystematic} reviews of clinical trials are at the foundation of evidence-based medicine and should represent comprehensive, high quality, and up to date syntheses of trial evidence. With the rapid growth of the scientific literature, identifying relevant evidence and keeping systematic reviews up to date is increasingly difficult. Studies examining the timing of systematic reviews suggest that reviews are updated on average every 5.5 years, though a substantial proportion should be updated within 2 years~\cite{Shojania07, Jaidee10, Garritty10, Peterson11}. Performing systematic reviews is time and resource intensive and even determining when a systematic review needs updating often requires completion of the searching and screening steps of the systematic review process. To facilitate this assessment, a number of tools and guidelines have been developed that aim to identify when new relevant research becomes available, or estimate the risk that the results of the systematic review may have substantially changed due to new evidence~\cite{Garritty10, Peterson11, Chung12, Pattanittum12, Takwoingi13, Ahmadzai13, Shekelle14, Garner16}.

These approaches rely on bibliographic databases, which are limited due to publication and reporting biases that affect the timing and completeness of the results~\cite{Chalmers09}. About half of all trials remain unpublished two years after trial completion, and of those that are published, around half have missing or changed outcomes~\cite{Dwan13, Dickersin90}. As a consequence, bibliographic databases may not provide a complete and timely source of relevant trial evidence for systematic reviews. As various policies and mandates are making prospective trial registration standard practice, clinical trial registries are an increasingly comprehensive and timely source of new research evidence, and in many cases may provide a more complete and less biased record than bibliographic databases~\cite{Bashir17}. However, the vast majority of methods aiming to support the identification of relevant studies for a systematic review operate over bibliographic databases rather than trial registries~\cite{OMaraEves15} and systematic reviews often fail to incorporate any clinical trial registries to identify relevant trials~\cite{Baudard17}. New methods for identifying relevant trials in clinical trial registries could help determine when systematic reviews need to be updated and support living systematic reviews and automated systematic review updates~\cite{Elliott14, Tsafnat14, Tsafnat13}.

Our aim was to evaluate a new method to partially automate the identification of trials that may be relevant for systematic review updates given the existing trials in a systematic review. This process could serve to signal when a systematic review becomes out of date, based on the amount and type of new evidence that is detected.

\section{Related Work}\label{sec.related}

A number of semi-automated methods have been proposed to identify relevant trials for inclusion in systematic reviews and improve the efficiency of the searching and screening processes~\cite{OMaraEves15, Cohen06, Aphinyanaphongs05}. The methods typically use the words or concepts included in the text of published articles to find similarities that are then used to distinguish relevant from irrelevant articles. Some work has also been done to directly extract information on populations, interventions, comparators, and outcomes~\cite{Kim16, Olorisade17}, which can then be used to match search queries. Several approaches have included the use of active learning~\cite{Miwa14, Wallace10}, while others have examined representations that use neural network based vector space models~\cite{Hashimoto16}. Far less work has been performed on identifying trials from the information stored in clinical trial registries or on linking clinical trial registries to bibliographic databases~\cite{Bashir17, Huser13, Huser12, Bashir16}. However, some methods have shown that it is possible to identify meaningful clusters of similar trials within registries~\cite{Hao14, Weng14, Boland13}, especially in relation to populations~\cite{He15}, and ClinicalTrials.gov data has been used in predicting black box warnings~\cite{Ma16}.

Matrix factorisation has the potential to support the identification of relevant trials for inclusion in systematic reviews. The approach has a long history of use in addressing problems in link prediction~\cite{Jamali10, Koren09, Menon11}, for example in building systems that recommend books, music, or new social connections to users. This process, commonly referred to as ``the item prediction problem'', aims to predict the presence or absence of links between the nodes of a graph where the vertices represent users and items, and edges that connect vertices are weighted according to preference scores. Matrix factorisation produces a mapping between users and items into a low-dimensional representation (latent factors) to model the user-item affinity in vector space.

Past work on the use of matrix factorisation for collaborative filtering focused on increasing prediction accuracy by including neighbourhood information~\cite{Bell07}. Later, Koren et al.~\cite{Koren08} proposed SVD++, a matrix factorisation approach that unified neighbourhood and latent factors. Guo et al.~\cite{Guo15} proposed TrustSVD, an extension of SVD++ that incorporates social trust information to help mitigate data sparsity and the cold start problem. TrustSVD includes factorisation of two matrices that share a same latent space—meaning a matrix of user-item preference scores and another matrix that defines trust information among users.

Our proposed matrix factorisation approach shares structural similarity with TrustSVD but applies different regularization methods to adapt to weighted links between systematic reviews and trial registrations. To the best of our knowledge, this is the first study to address the problem of recommending trial registrations for systematic review updates using matrix factorisation with a shared latent space.

\section{Materials and Methods}\label{sec.method}
\subsection{Study data}\label{sec.studydata}

\begin{figure}[!h]\centering
\includegraphics[width=3.6in,keepaspectratio=true,origin=c]{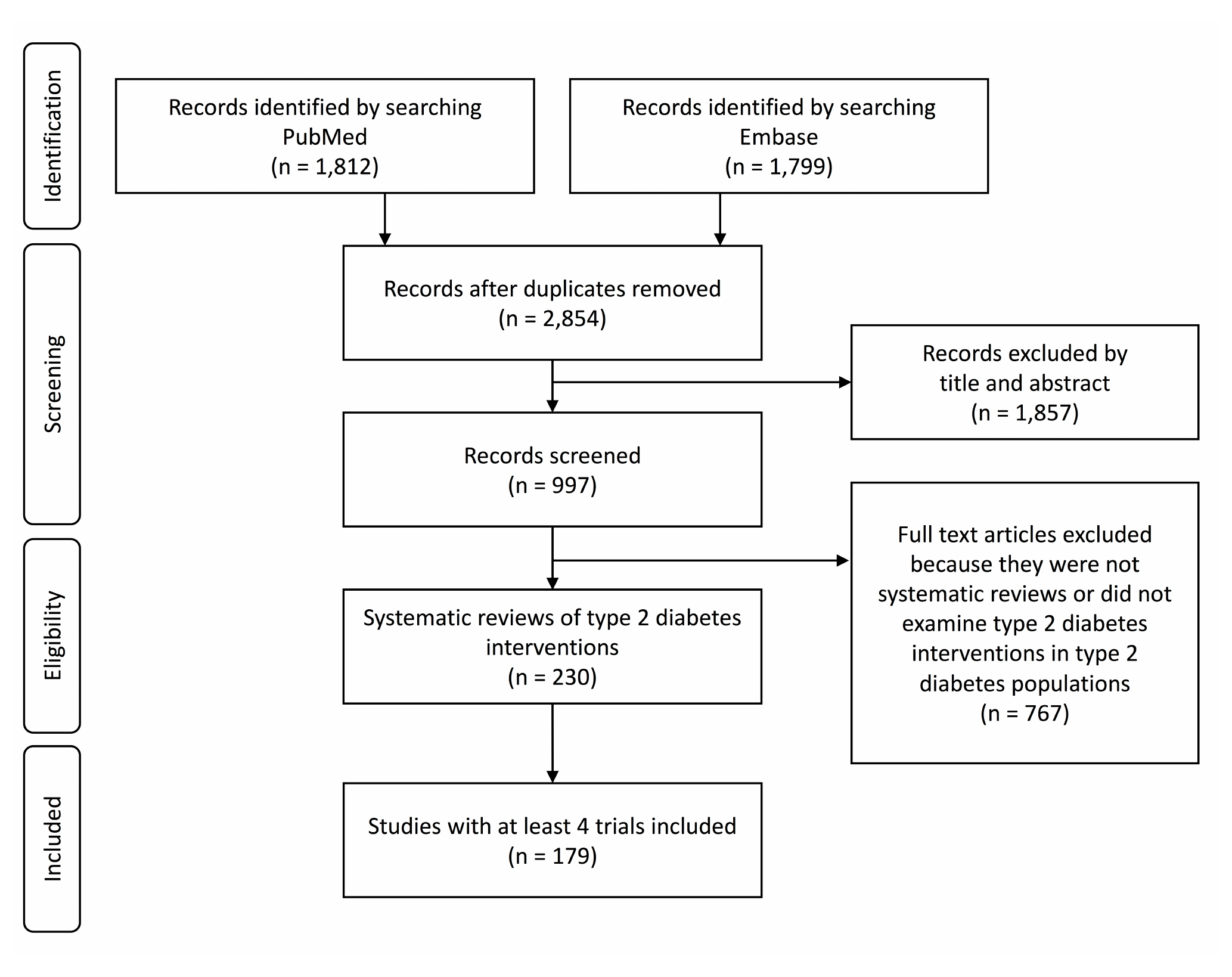}
\caption{From 2,854 unique articles identified in the search, 230 systematic reviews of type 2 diabetes were identified and 179 were included in the experiments.}
\label{fig:prisma}
\end{figure}

We searched PubMed and Embase for systematic reviews of drugs used to treat type 2 diabetes. The final search was performed on 26 March 2017, using a search strategy that included terms for type 2 diabetes, type 2 diabetes interventions, and publication type information. Systematic reviews were considered for inclusion in the experiments if they were focused on populations with type 2 diabetes and included at least one meta-analysis (Figure~\ref{fig:prisma}). Reviews that summarised or synthesised meta-analyses from other reviews were excluded, as were reviews that did not include full details of a search strategy and did not include at least one meta-analysis of a safety or efficacy outcome. Reviews that also included other populations were included only if they had sub-group analyses that were specific to type 2 diabetes.

After excluding systematic reviews that did not have links to at least 4 trial registrations, we had 179 systematic reviews with 4,447 links to 537 unique trial registrations. For each systematic review, we ordered the registrations by trial completion date to use older trials for training and newer trials for testing, simulating the way new trials might be added to existing systematic reviews. Trial registrations had a median of 14 links (ranging from 4 to 147) to the systematic reviews. 

We accessed ClinicalTrials.gov on May 16, 2017 to create a static local version of information from 128,392 completed trials in order to produce consistent results across all experiments. Information on the brief and official titles, detailed description, inclusion criteria, and intervention names were extracted for use in the experiments.

To test the methods in a general scenario, we additionally identified 17 systematic reviews that were published in the Cochrane Database of Systematic Reviews on any topic. To be included in the experiments, they must have either had an update published with a new search performed before 25 April 2017, or have listed at least three ongoing studies in ClinicalTrials.gov in the published review. We then used the set of included trials as the training set, and the new trials added to the update (or listed as ongoing trials) as the test set. There were 72 unique trials in the original reviews and 69 unique trials in the updates and lists of ongoing relevant trials. The number of trial registrations per systematic review was between 1 and 14 in the original versions and between 1 and 13 in the updates.

\subsection{Feature representations}\label{sec.featurerepresentation}

Each of the trial registrations was considered as a single document and treated as a bag of words (order was not considered). Each word in the text was converted to lowercase, standard stop words were excluded, the Porter stemmer~\cite{Porter97} was applied, and words present in fewer than five trial registrations were excluded. Each trial registration was represented by the set of extracted words from the text after the pre-processing step.

We used multiple feature representations for each trial registration. First, a binary vector representation was created to indicate whether a word feature is present or absent in the text of a trial registration. Secondly, a word frequency vector representation was used to capture the number of times a word feature appears in the text of a trial registration. Lastly, we used a term frequency-inverse document frequency (tf-idf) vector representation, where the entry of the vector is the tf-idf score of a word feature. In the experiments that follow, we refer to these three feature representations as full-dimension feature representations.

We used Principal Component Analysis (PCA) and Latent Dirichlet Allocation (LDA) to reduce the dimensionality of the feature spaces. The PCA technique transforms data into a lower-dimensional space based on several linearly uncorrelated variables by maximising the variance. We used the implementation of incremental PCA from scikit-learn with its default settings, and tested it with 20, 50, 100, 200, 300, and 400 components on the tf-idf and word frequency vector representations. The LDA method is a technique that was first introduced to model topics across a set of documents~\cite{Crain12, blei03}. It uses co-occurrence of words to find latent structures, or topics, in a corpus of documents and a topic is represented by a distribution of word probabilities for that topic. Using LDA, a trial registration is represented by a distribution of topics, where the number of topics is set by parameter. For LDA, we used the gensim implementation with standard settings~\cite{lu11}, and tested it with 20, 50, 100, 200, 300 and 400 topics. We refer the feature representations using PCA and LDA as reduced-dimension feature representations.

\begin{figure}[!h]\centering
\includegraphics[width=3.2in,keepaspectratio=true,origin=c]{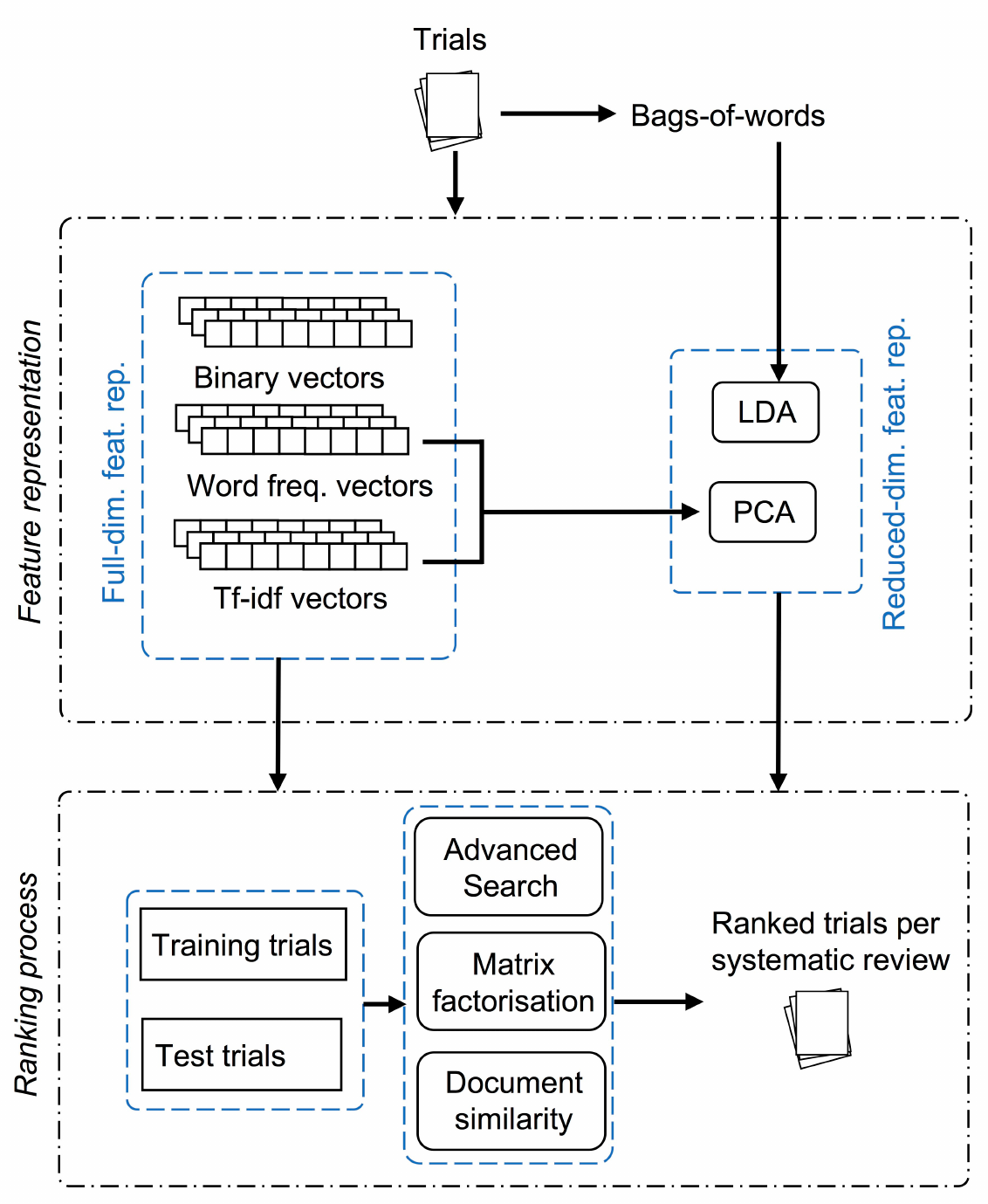}
\caption{High-level view of ranking process for trial inclusion in systematic review updates.}
\label{fig:process}
\end{figure}

Given a systematic review, one approach to identify relevant trial registrations is to rank all trial registrations according to their similarity with the set of trial registrations for trials included in the review. To determine similarity, we used three document similarity measures: cosine similarity, Euclidean distance, and squared Euclidean distance. These measures were used as a baseline approach against which the matrix factorisation approach was tested. The overall process is illustrated in Figure\ref{fig:process}.

\subsection{Matrix factorisation with a shared latent space}\label{sec.matrixfactorization}

Matrix factorisation decomposes a matrix into products of matrices. In our approach, we combine two available sources of information: the text from each trial registration and the links between the trial registrations and the systematic reviews in which they were included. The approach integrates both sources of information via a shared latent space in the learning process to rank all other trial registrations relative to the systematic reviews.

Given $U$ trial registrations and $V$ systematic reviews, where each trial registration is represented by an $J$-dimensional feature vector extracted from its text and an $V$-dimensional binary vector to denote the absence or present of a link between a trial registration and a systematic review, we created two matrices: a matrix of trial registrations-features, $R$ $\in$ $\mathbb{R}^{U \times J}$; and a matrix of trial registrations-systematic reviews, $T$ $\in$ \{0,1\}$^{U \times V}$. Given $K$ latent factors (where $K$ $\ll$ $J$ and $K$ $\ll$ $V$), matrices $R$ and $T$ are then decomposed into the following:

\begin{align}
&R \approx P Q^\top, P \in \mathbb{R}^{U \times K}, Q \in \mathbb{R}^{J \times K} \nonumber\\
&T \approx P W^\top, P \in \mathbb{R}^{U \times K}, W \in \mathbb{R}^{V \times K} \nonumber
\end{align}

\begin{figure}[!h]\centering
\includegraphics[width=3.5in,keepaspectratio=true,origin=c]{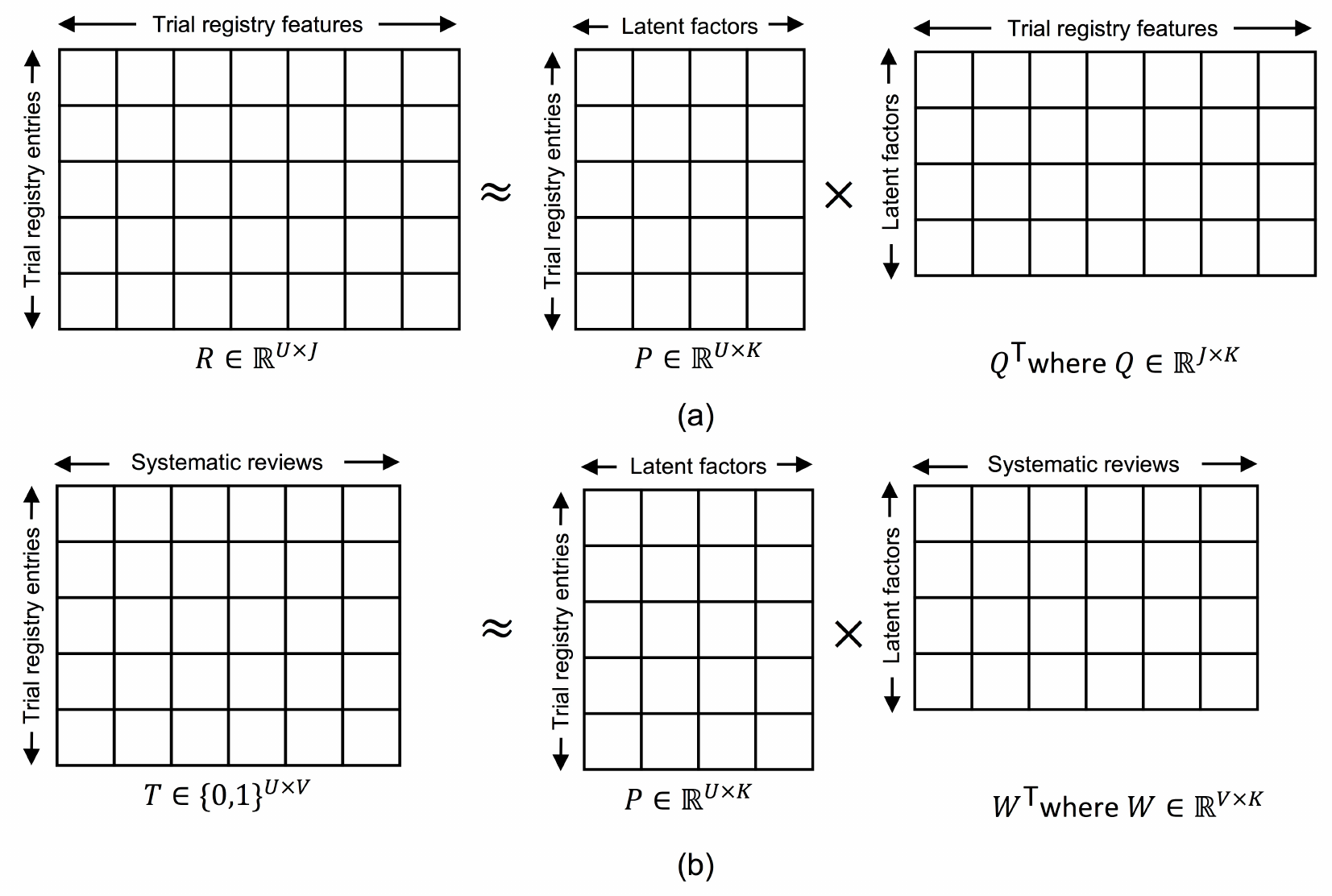}
\caption{The illustration of matrix factorisation for (a) the matrix of trial registry entries-features; and (b) the matrix of trial registries-systematic reviews.}
\label{fig.factorization}
\end{figure}

In these equations, $P$ is a matrix of trial registration-latent factors, $Q$ is a matrix of trial registration feature-latent factors and $W$ is a matrix of systematic review-latent factors (Figure~\ref{fig.factorization}).

Note that the matrix $P$ is shared and used to connect both $R$ and $T$ in the decomposition process. In other words, the matrix $P$ will describe how trial registrations are associated with systematic reviews based on the features of trial registrations. The goal of matrix factorisation is then to learn matrices $P$, $Q$, and $W$. Once the matrices are learnt, their respective dot products approximate values in matrices $R$ and $T$; \ie\ $\hat{R}$ $\approx$ $P$$Q^\top$ and $\hat{T}$ $\approx$ $P$$W^\top$. For the unknown values in matrix $T$ where link information between trial registrations and systematic reviews is not present, the respective approximated values in matrix $\hat{T}$ are used as a measure of similarity that can then be used to rank unknown trial registrations for each systematic review.

Because the matrix of trial registrations-features, $R$ $\in$ $\mathbb{R}^{U \times J}$, is decomposed into a matrix of trial registration-latent factor $P$ $\in$ $\mathbb{R}^{U \times K}$ and a matrix of trial registration feature-latent factor $Q$ $\in$ $\mathbb{R}^{J \times K}$, the loss function that we need to minimise is as follows~\cite{Menon11, Guo15}: 

\begin{eqnarray}\label{eq.1}
\mathscr{L} = \frac{1}{2}\sum\limits_{u}\sum\limits_{j}(\hat{r}_{uj} - r_{uj})^2 + \frac{\lambda}{2}(\sum\limits_{u}\|p_u\|^2_F + \sum\limits_{j}\|q_j\|^2_F) \nonumber\\
\end{eqnarray}

\noindent where $|| \cdot ||$ represents the Frobenius norm, $\lambda$ is the regularization parameter, $r_{uj}$ is the entry of matrix $R$ in row $u$ and column $j$, and $\hat{r}_{uj}=p_uq_j^{\top}$. $p_u$ and $q_j$ represent $K$-dimensional vector of trial registration $u$ and its feature $j$ respectively.

In a similar way, the matrix of (trial registrations-systematic reviews), $T$ $\in$ \{0,1\}$^{U \times V}$, is decomposed into a matrix of (trial registration-latent factor) $P$ $\in$ $\mathbb{R}^{U \times K}$ and a matrix of (systematic review-latent factor) $W$ $\in$ $\mathbb{R}^{V \times K}$, and the loss function to minimise is as follows: 

\begin{eqnarray}\label{eq.2}
\mathscr{L} = \frac{1}{2}\sum\limits_{u}\sum\limits_{v \in T_u}(\hat{t}_{uv} - t_{uv})^2 + \frac{\lambda}{2}(\sum\limits_{u}\|p_u\|^2_F + \sum\limits_{v}\|w_v\|^2_F) \nonumber\\
\end{eqnarray}

\noindent where $T_u$ represents a set of systematic reviews that a trial registration belongs to, $t_{uv}$ is the entry of matrix $T$ in row $u$ and column $v$, and $\hat{t}_{uv}=p_uw_v^{\top}$. $p_u$ and $w_v$ represent $K$-dimensional vector of trial registration $u$ and systematic review $v$ respectively.

Because matrix  is shared in Equation~\ref{eq.1} and~\ref{eq.2}, the new objective function to minimise is then given as follows:

\begin{eqnarray}\label{eq.3}
\mathscr{L} & = & \frac{1}{2}\sum\limits_{u}\sum\limits_{j}(\hat{r}_{uj} - r_{uj})^2 + \frac{\lambda_t}{2}\sum\limits_{u}\sum\limits_{v \in T_u}(\hat{t}_{uv} - t_{uv})^2 \nonumber\\
                   & + & \frac{\lambda}{2}(\sum\limits_{u}\|p_u\|^2_F + \sum\limits_{j}\|q_j\|^2_F + \sum\limits_{v}\|w_v\|^2_F)
\end{eqnarray}

We performed gradient descent on $p_u$, $q_j$ and $w_v$ based on the final loss function in Equation~\ref{eq.3} described in the following equations: 

\begin{eqnarray}\label{eq.4}
\frac{\partial \mathscr{L}}{\partial p_u} & = & \sum\limits_{u}\sum\limits_{j} e_{uj}q_j^{\top} + \lambda_t \sum\limits_{u}\sum\limits_{v \in T_u}e_{uv} w_v^{\top} + \lambda \sum\limits_{u} p_u \nonumber\\
\frac{\partial \mathscr{L}}{\partial q_j} & = & \sum\limits_{u}\sum\limits_{j} e_{uj}p_u + \lambda\sum\limits_{j} q_j \nonumber\\
\frac{\partial \mathscr{L}}{\partial w_v} & = & \lambda_t\sum\limits_{u}\sum\limits_{v} e_{uv}p_u + \lambda\sum\limits_{v} w_v 
\end{eqnarray}

\noindent where $e_{uj}=\hat{r}_{uj}-r_{uj}$ represents the feature prediction error for trial registration $u$ on feature $j$, and $e_{uv}=\hat{t}_{uv}-t_{uv}$ represents the link prediction error for trial registration $u$ and systematic review $v$.

We used 5,000 as our maximum number of iterations for the learning process. We calculated root-mean-square error (RMSE) in each iteration during the learning process and returned the results from the learning process when the RMSE was the lowest. We implemented our matrix factorisation approach in \textit{Cython}. We performed the experiments with $\lambda$=0.001, $\lambda_t$=0.01, and the number of latent factors was set to 5, 10, 20, 30, 40, and 50. The source code of the proposed matrix factorisation approach has been made available via a repository (https://github.com/dsurian/matfac).

\subsection{Experiments and outcome performance measures}\label{sec.measure}

For each systematic review, we split the set of links between included trials and systematic reviews into training and test sets with a minimum of 3 links included in the training for each systematic review, and kept this set constant for all experiments. Because systematic reviews had a different number of included trials, the training set included between 3 and 89 links (a median of 9) per systematic review, and a test set with between 1 and 58 links (a median of 5) per systematic review. In the Cochrane systematic reviews, the training set comprised the set of trials included in the systematic reviews (between 1 and 14 links) and the test set comprised the set of trials included in the update of the systematic reviews (between 1 and 13 links).

In each experiment and for each systematic review we produced a ranking for all 128,392 completed ClinicalTrials.gov registrations, and identified where the test links for that systematic review were found in the rankings. For each experiment, the aim was to rank the links from the test set as high as possible within the full set of candidates. The links corresponding to the training set were not included in the ranking and did not contribute to the performance measures.

We calculated the median rank of the trials in the test set by aggregating the ranks of all of the links in the test set and taking the median (i.e. each trial-systematic review link contributed one score and the median rank was calculated from all of its scores). Using this approach, if we were to assign a random rank to all of the trials in the test set, the median rank would converge to approximately 64,196.

We evaluated the proportion of links in the test set that were included in the first 100 candidates in each of the ranked lists (recall@100), and produced a function for the general version of the same measure (recall@N). The recall@N function can be used to determine the number of candidates that need to be examined per systematic review to identify a given proportion of relevant trial registrations. We also included the work saved over sampling (WSS)~\cite{Cohen06} which shows the percent reduction in screening process to retrieve the relevant trials. We measured WSS at recall 95\% (WSS@95\%) by calculating the proportion of registrations that did not need to be screened after screening the ranked candidate registrations and identifying at least 95\% of the registrations in the test set.

To evaluate the matrix factorisation and document similarity approaches against an appropriate baseline, we undertook a manual search of ClinicalTrials.gov using the Advanced Search function, which ranks trials based on a relevance score. One investigator (AD) replicated the search strategies of the reviews in the format of population, intervention, and outcome (as required), and the returned search results were used as a ranked set. These results were evaluated using the same outcome measures.

\section{Results}\label{sec.results}

\begin{figure}[!h]\centering
\includegraphics[width=3.5in,keepaspectratio=true,origin=c]{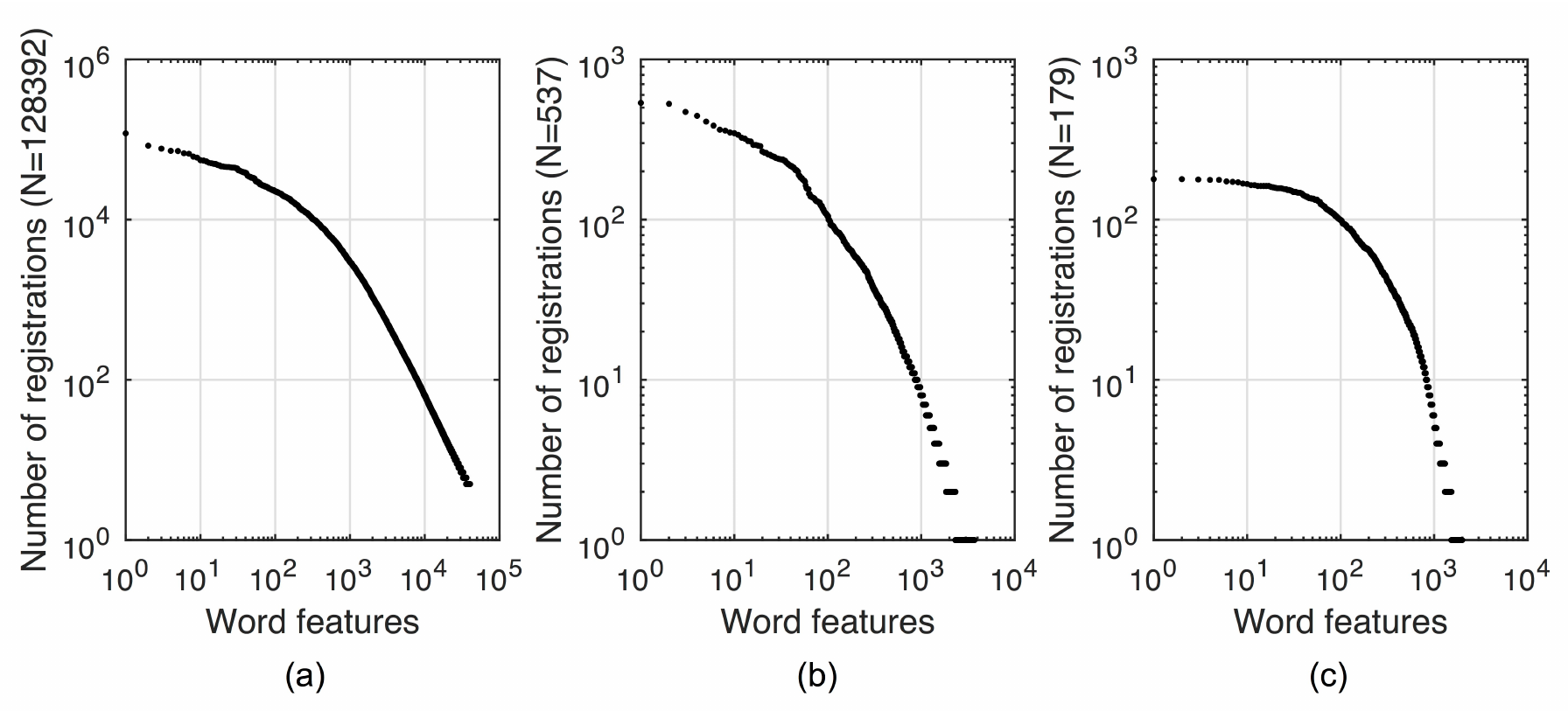}
\caption{The distribution of the frequency with which each word feature was found: (a) at least once in any trial registration; (b) at least once in the 537 type 2 diabetes trial registrations; and (c) shared across the training and test sets for each of the 179 systematic reviews.}
\label{fig.worddistribution}
\end{figure}

We identified 40,049 unique words in the vocabulary extracted from the processed text of trial registrations. For the binary and word frequency vector representations, the number of features that were shared between training and test sets was 2,367, comprising 5.9\% of the 40,049 words (Figure~\ref{fig.worddistribution}).

For the document similarity methods, the full set of word features with tf-idf weights consistently produced higher levels of performance (Table~\ref{tb.allresults}). In the best performing methods, half of the correct links could be identified by checking 138 candidates (i.e., median rank of 138), and 42.8\% of correct links could be identified by checking 100 candidates. Compared to the manual search, the results from the document similarity method were 9.2\% lower for the median rank and 33.3\% higher for recall@100.

The best performing matrix factorisation approach produced a recall@100 of 60.9\% (and a median rank of 59), which was achieved with the LDA feature representation using 200 topics, and 50 latent factors. The median rank of the best-performing matrix factorisation approach was 57.2\% lower and the recall@100 was 42.3\% higher than the best performing document similarity approach, and both methods gave a similar WSS@95\%. Compared to the manual search, the results from the matrix factorisation were 61.2\% lower for the median rank and 89.7\% higher for the recall@100. The performance of matrix factorisation using the full-dimension feature representations was consistently lower than the performance of document similarity.

\begin{table*}[!htb]\centering\small
\captionsetup{labelsep=newline,justification=centering}
\renewcommand{\arraystretch}{1.3}
\caption{The results from the manual search compared to the highest performing document similarity methods and matrix factorisation.}
\label{tb.allresults}
\begin{tabular}{lrrr}\hline
 & \textbf{WSS@95\%} & \textbf{Median Rank} & \textbf{Recall@100} \\ \hline\textbf{Manual Search} & \textbf{97.0\%} & \textbf{152} & \textbf{32.1\%} \\ \hline
\textbf{Document Similarity} & & & \\
\hspace{0.5cm}\textbf{Full tf-idf weights} & & & \\
\hspace{1.0cm}Euclidean distance & 99.5\% & 139 & 42.6\% \\
\hspace{1.0cm}Squared Euclidean distance & \textbf{99.5\%} & \textbf{138} & \textbf{42.8\%} \\
\hspace{1.0cm}Cosine similarity & 99.5\% & 154 & 41.7\% \\
\hspace{0.5cm}\textbf{PCA (400 components), tf-idf weights} & & & \\
\hspace{1.0cm}Euclidean distance & 95.5\% & 329.5 & 29.4\% \\
\hspace{1.0cm}Squared Euclidean distance & 95.4\% & 341 & 27.4\% \\
\hspace{1.0cm}Cosine similarity & 99.1\% & 195 & 37.0\% \\
\hspace{0.5cm}\textbf{LDA (400 topics)} & & & \\
\hspace{1.0cm}Euclidean distance & 94.0\% & 363 & 23.4\% \\
\hspace{1.0cm}Squared Euclidean distance & 93.5\% & 409.5 & 22.2\% \\
\hspace{1.0cm}Cosine similarity & 98.0\% & 433 & 20.4\% \\ \hline
\textbf{Matrix Factorisation} & & & \\
\hspace{0.5cm}\textbf{PCA (300 components), tf-idf weights} & & & \\
\hspace{1.0cm}5 latent factors & 99.4\% & 76 & 57.6\% \\
\hspace{1.0cm}10 latent factors & 99.3\% & 75 & 57.6\% \\
\hspace{1.0cm}20 latent factors & 99.4\% & 75 & 57.7\% \\
\hspace{1.0cm}30 latent factors & 99.4\% & 76 & 58.0\% \\
\hspace{1.0cm}40 latent factors & 99.3\% & 74 & 58.1\% \\
\hspace{1.0cm}50 latent factors & 99.4\% & 74 & 58.6\% \\
\hspace{0.5cm}\textbf{LDA (400 topics)} & & & \\
\hspace{1.0cm}5 latent factors & 96.2\% & 78 & 57.1\% \\
\hspace{1.0cm}10 latent factors & 99.0\% & 75.5 & 57.6\% \\
\hspace{1.0cm}20 latent factors & 99.2\% & 77 & 57.4\% \\
\hspace{1.0cm}30 latent factors & 99.2\% & 74 & 57.9\% \\
\hspace{1.0cm}40 latent factors & 99.2\% & 59 & 60.7\% \\
\hspace{1.0cm}50 latent factors & \textbf{99.2\%} & \textbf{59} & \textbf{60.9}\% \\ \hline
\end{tabular}
\end{table*}

\begin{figure*}[!h]\centering
\includegraphics[width=5.5in,keepaspectratio=true,origin=c]{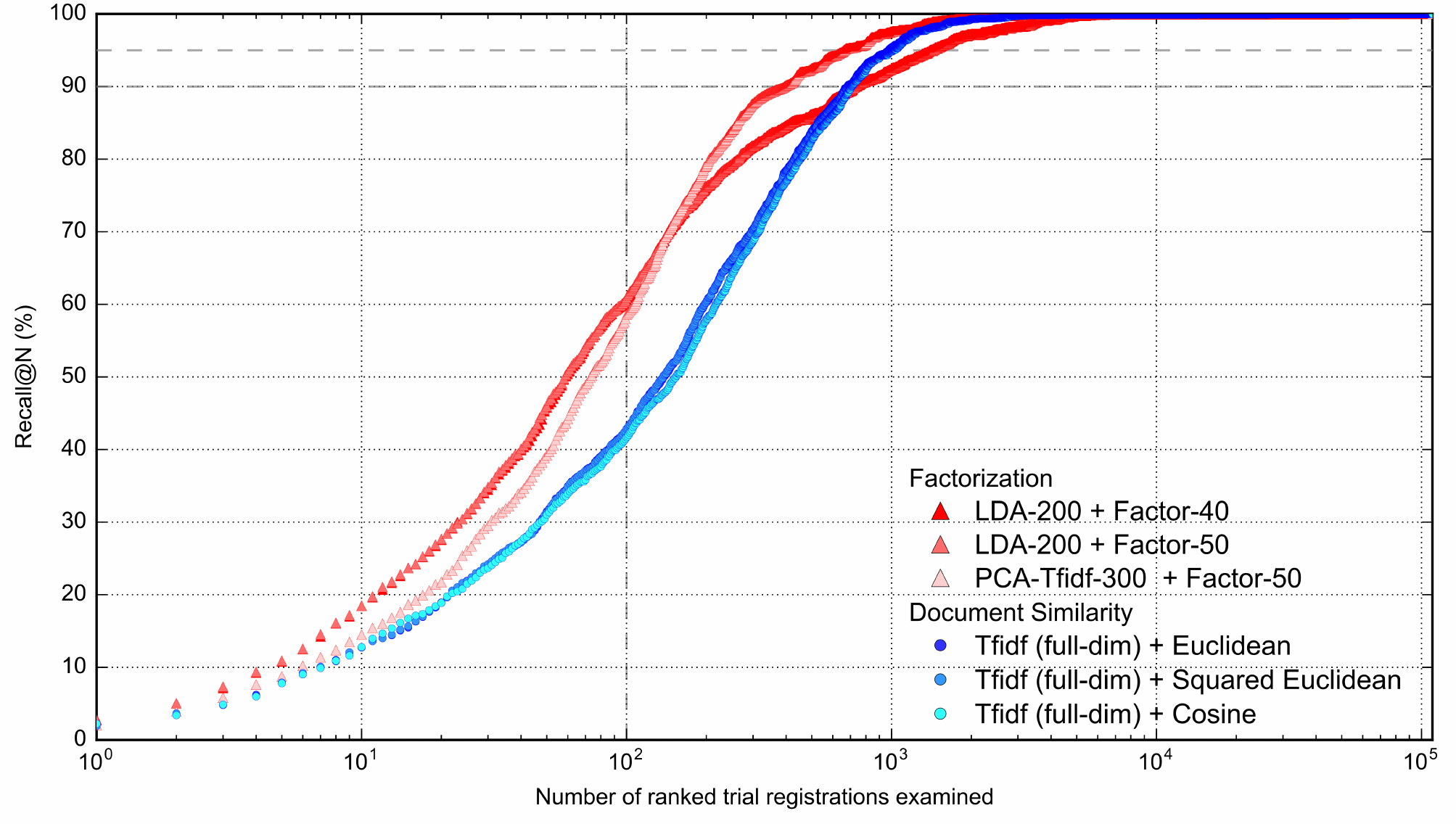}
\caption{Recall@N for various number of candidates that were needed to be checked before the trial registrations included in the type 2 diabetes systematic reviews were found among the set of 128,392 trial registrations. The horizontal dashed-lines mark the positions where recall 90\% and 95\% were achieved.}
\label{fig.recallfactor}
\end{figure*}

Using the matrix factorisation approach, a user would need to examine the top 400 candidates to find 90\% of the test set, and after examining the top 700 candidates, approximately 95\% of the test set would have been found (Figure~\ref{fig.recallfactor}).

\subsection{Tests in a general scenario}\label{sec.practical}

The document similarity baseline gave the best performance compared to the matrix factorisation approach in the general scenario, which had fewer training examples. The best-performing document similarity method produced a median rank of 67 (recall@100: 62.9\%) using all word features and tf-idf weights and cosine similarity, while the best-performing matrix factorisation approach produced a median rank of 5,124 (recall@100: 1.4\%) using LDA with 20 topics and 30 latent factors. 

\begin{figure}[!h]\centering
\includegraphics[width=3.5in,keepaspectratio=true,origin=c]{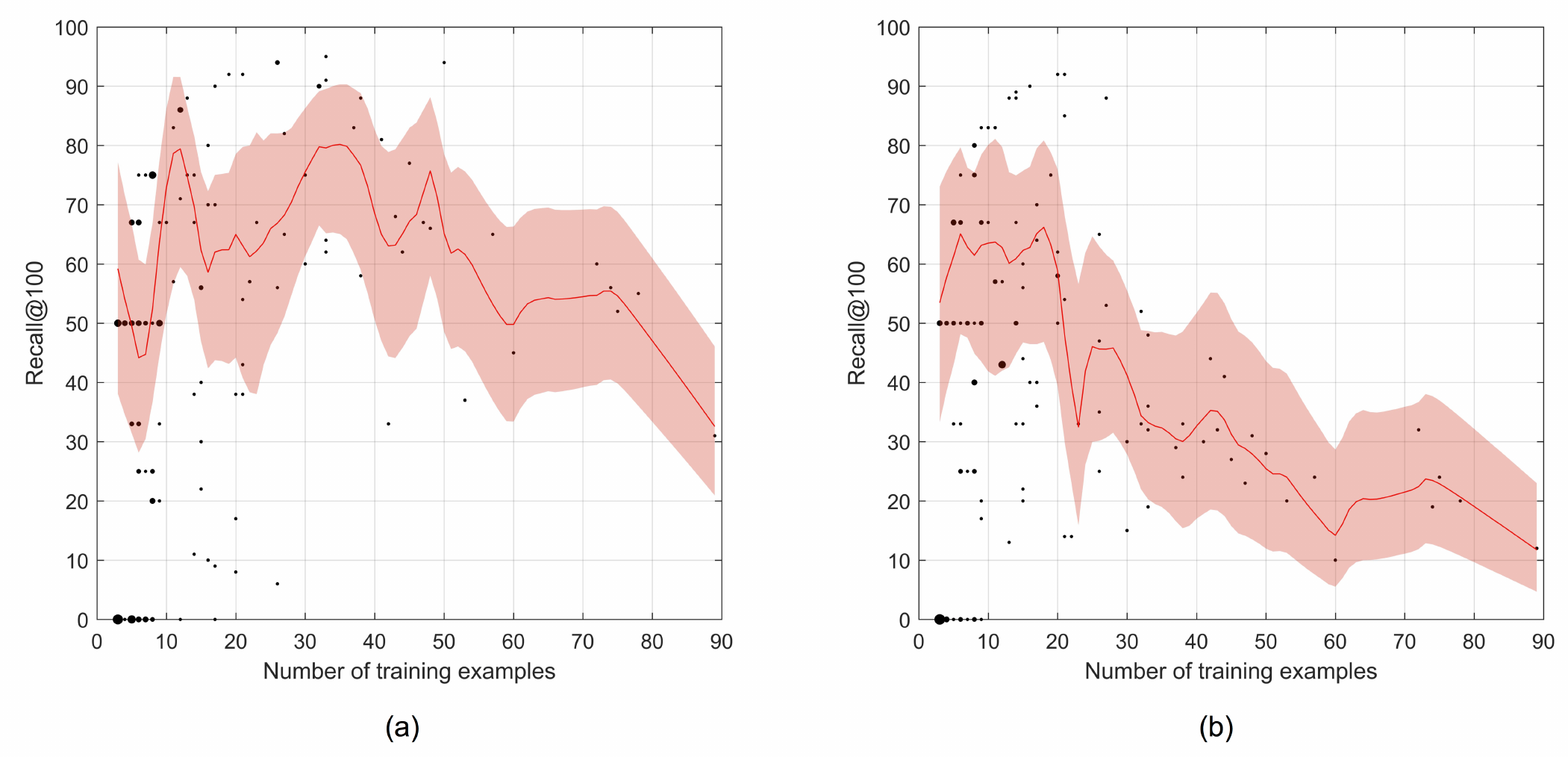}
\caption{Changes in the recall@100 relative to the number of training examples for (a) the highest-performing matrix factorisation method; and (b) the highest performing document similarity method when applied to the 179 type 2 diabetes reviews. Each systematic review contributes to the size of the black dots, and the moving average (range of 7) and its exact 95\% confidence interval are given in red as a guide.}
\label{fig.numtraineffect}
\end{figure}

The performance of the matrix factorisation method was lower in the general scenario than in the type 2 diabetes examples, and lower than the document similarity methods. To investigate possible reasons, we looked at how the results varied with the number of training examples in the type 2 diabetes systematic reviews (Figure~\ref{fig.numtraineffect}). The results indicate that the document similarity method degrades as the number of examples decreases, while the matrix factorisation tends to improve or maintain its performance as the number of training examples increases (see Discussion and Figure~\ref{fig.tsne}).

\section{Discussion}\label{sec.discussion}

Our results indicate that a shared latent space matrix factorisation method can support the identification of trial registrations that are relevant to a systematic review. Both the matrix factorisation and document similarity methods could be used together as part of a process to signal when a systematic review is likely to be out of date.

\subsection{Comparisons with existing research}\label{sec.comparison}

This is the first study to evaluate matrix factorisation for the purpose of finding new relevant trials from ClinicalTrials.gov for systematic review updates. In terms of manual screening requirements, our findings show that the performance of our methods compares favourably to similar methods that operate over bibliographic databases. While the results of studies testing semi-automated screening of published articles for inclusion in systematic reviews are not directly comparable, standard tools in this area use active learning to avoid screening 80\% of articles to reach 95\% recall~\cite{Cohen06, Wallace10, Hashimoto16}. In our experiments, we were able to reach 95\% recall after screening approximately 700 candidates (avoiding 99.5\% of the candidate registrations).

A recent study using citation links trained on 23 systematic reviews found that to reach 75\% recall, the precision dropped to 3.6\%~\cite{Belter17}, retrieving 300 correct citations after screening 8,298 articles. For our approach trained on 179 systematic reviews, 75\% recall was reached after screening fewer than 200 trial registrations.

In another recent example, investigators used generalized linear models and gradient boosting machines on 3 systematic reviews, and needed to screen between 192 and 2,112 articles to reach 96\% recall~\cite{Shekelle17}. The approach is similar to ours in that it requires knowing which studies were included in existing systematic reviews. However, the approach appears to have used between 6,502 and 41,066 negative training examples and between 55 and 356 positive training examples. By comparison, our approach demonstrated that investigators could screen a similar number of trial registrations to reach the same level of recall, but using fewer positive training examples with no negative training examples. This was possible because the matrix factorisation approach we proposed takes advantage of the latent structure of the set of other training examples and is therefore expected to improve with time as more training examples become available. 

\subsection{Implications and recommendations}\label{sec.implication}

As policies and practice around the prospective registration and structured results reporting of clinical trials continue to expand, use of trial registries in the production of systematic reviews will likely increase. An important benefit will include earlier identification of when the evidence summarised in a systematic review has become out of date. Tools that automate the identification of relevant trials or dramatically reduce the human workload required in trial selection, could be applied to provide rapid estimates of how much of the available evidence is covered in a published systematic review, thus signalling the need for an update. If an update cannot be produced in a specified time, the systematic review could be flagged as potentially out of date with reference to the new research, in order to inform clinicians and consumers using systematic reviews in their care decisions.

\begin{figure*}[!h]\centering
\includegraphics[width=5in,keepaspectratio=true,origin=c]{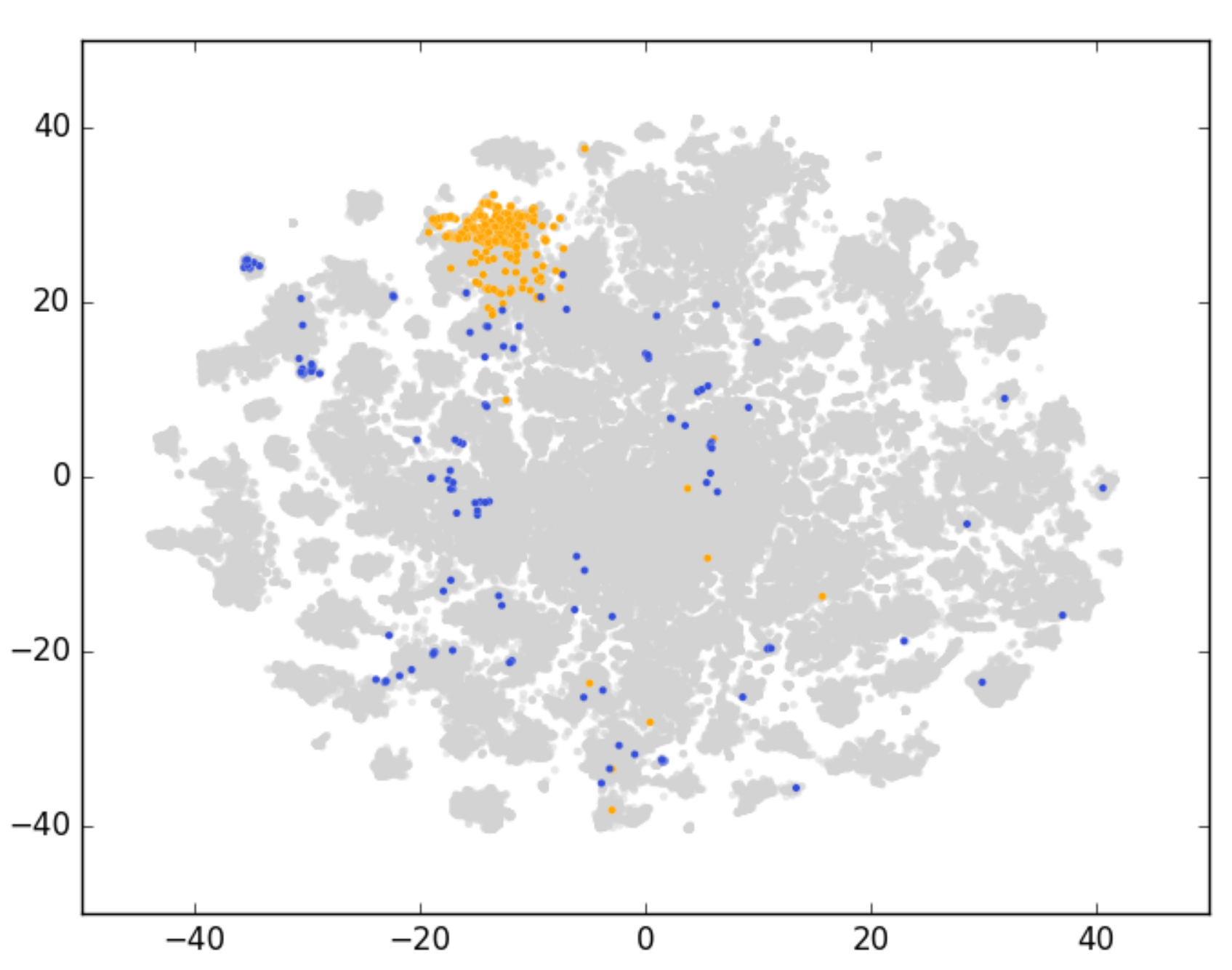}
\caption{The t-SNE visualisation of the trial registration space constructed from the PCA feature representation. The 537 unique trial registrations included in the type 2 diabetes reviews (orange), and the 141 unique trial registrations in the Cochrane reviews (blue) are highlighted among the set of 128,392 trial registrations (grey).}
\label{fig.tsne}
\end{figure*}

Our results suggest that the matrix factorisation approach works best when there are more trials for training from similar systematic reviews. While the document similarity approach only uses included trials from one systematic review to find close examples of other trial registrations, the matrix factorisation method uses information from other nearby systematic reviews to learn how to rank registrations. Figure~\ref{fig.tsne} is a t-SNE visualisation~\cite{Maaten08} of the 128,392 completed trial registrations, illustrating the differences between the 537 trials included in the 179 type 2 diabetes systematic reviews (orange) and the 141 trials included in the 17 Cochrane systematic reviews (blue). In the figure, the position is determined by mapping the feature space into two dimensions such that similar trials are located close together. The matrix factorisation is better able to learn the features that reproduce the known links available in the orange region of this space, and struggles to learn the important features for the Cochrane reviews because the known links are more sparsely distributed across the 128,392 completed trial registrations from ClinicalTrials.gov.

The results of the experiments suggest that the document similarity and matrix factorisation approaches can replace the need to undertake a search of trial registrations as well as improve the efficiency of screening for relevant trials. This suggests that the two could be used as complementary solutions as part of a pipeline of methods for automatically identifying trials that may be relevant to the update of a systematic review, or used to support updates of living systematic reviews~\cite{Elliott14}.

\subsection{Limitations}\label{sec.limitation}

The study has several limitations. First, the set of systematic reviews used in the training and validation of the methods are related to drug interventions in type 2 diabetes--an area with a substantial number of new drugs and a large volume of new trials. This means that the results may not generalise to clinical application domains with fewer trials available~\cite{Thomas17}, or where trial descriptions are more heterogeneous. Second, our results may underestimate the performance of both the baseline document similarity approach and the matrix factorisation approach because we only used known links to test the performance, and other highly--ranked candidates--especially those completed since the systematic reviews were published--may also have been relevant but not included in the systematic review. 

\section{Conclusion}\label{sec.conclusion}

The use of clinical trial registries to monitor which and when systematic reviews require updating has been limited by the extensive manual processes required to identify relevant trial registrations. To date, semi-automated article screening methods used in bibliographic databases have not been extended to trial registries. We found that a matrix factorisation method can be used to rank trial registrations such that 75\% of relevant trial registrations will appear within the top 200 candidates. The matrix factorisation approach for identifying trials relevant to a systematic review update is likely to be most useful in practice if implemented in a pipeline or in combination with other methods designed for scenarios where no or few relevant trial registrations are known.


\ifCLASSOPTIONcompsoc
  \section*{Acknowledgments}
\else
  \section*{Acknowledgment}
\fi

FB and AD report funding from the Agency for Healthcare Research and Quality (R03HS024798).\\

\ifCLASSOPTIONcaptionsoff
  \newpage
\fi

\bibliographystyle{IEEEtran}
\bibliography{IEEEabrv,refbib}

\section{Appendix}
The search strategy for identifying systematic reviews of type 2 diabetes for PubMed (and translated to equivalent terms for Embase):
\\

\footnotesize{((type 2[Title/Abstract] OR type II[Title/Abstract] OR adult[Title/Abstract] OR slow[Title/Abstract]) 
AND (diabete*[Title/Abstract] OR diabetic*[Title/Abstract]))
AND ("meta analysis"[Publication Type] OR "review"[Publication Type] OR systematic review[Title/Abstract] OR meta analysis[Title/Abstract])
AND (metformin[Title/Abstract] OR glucophage[Title/Abstract] OR dipeptidyl-peptidase iv inhibitors[Title/Abstract] OR *gliptin[Title/Abstract] OR Januvia[Title/Abstract] OR glucagon-like peptide[Title/Abstract] OR Galvus[Title/Abstract] OR exenatide[Title/Abstract] OR Trajenta[Title/Abstract] OR Byetta[Title/Abstract] OR Onglyza[Title/Abstract] OR Bydureon[Title/Abstract] OR liraglutide[Title/Abstract] OR Victoza[Title/Abstract] OR lixisenatide[Title/Abstract] OR Lyxumia[Title/Abstract] OR thiazolidinedione*[Title/Abstract] OR glitazone*[Title/Abstract] OR *glitazone[Title/Abstract] OR Avandia[Title/Abstract) OR sulfonylurea*[Title/Abstract] OR sulphonylurea*[Title/Abstract] OR tolbutamide[Title/Abstract] OR glibenclamide[Title/Abstract] OR glipizide[Title/Abstract] OR Minidiab[Title/Abstract] OR glimepiride[Title/Abstract] OR Amaryl[Title/Abstract] OR gliclazide[Title/Abstract] OR Diamicron[Title/Abstract])}

\end{document}